\documentclass[preprint,aps]{revtex4}

\usepackage{graphicx}
\usepackage{dcolumn}
\usepackage{bm}

\begin{document}

\preprint{APS/123-QED}

\title{Evolution of electromagnetic and Dirac perturbations around a black hole in Ho\v{r}ava gravity}

\author{Nijo Varghese}
\author{V C Kuriakose}%
 \email{nijovarghese@cusat.ac.in, vck@cusat.ac.in}
\affiliation{%
Department of Physics, Cochin University of Science and Technology,
Cochin - 682 022, Kerala, India.
}%

\date{\today}

\begin{abstract}
The evolution of electromagnetic and massless Dirac
perturbations in the space-time geometry of Kehagias-Sfetsos(KS)
black hole in the deformed Ho\v{r}ava-Lifshitz(HL) gravity is
investigated and the associated quasinormal modes(QNMs) are
evaluated from time domain integration data. We find a considerable
deviation in the nature of field evolution in HL theory from that in
the Schwarzschild space-time and QNM region extends over a longer
time in HL theory before the power-law tail decay begins. The
dependence of the field evolution on the HL parameter $\alpha$ is
also studied. In the time domain picture we find that the length of
QNM region increases with $\alpha$. But the late time decay of field
follows the same power-law tail behavior as in the case of
Schwarzschild black hole.
\end{abstract}

\pacs{Valid PACS appear here}

\maketitle

\section{Introduction}
Recently a renormalizable theory of gravity in 3+1 dimensions was
proposed by Ho\v{r}ava, inspired from the Lifshitz theory in solid
state physics, now known as Ho\v{r}ava-Lifshitz(HL)
theory\cite{horava1,horava2,horava3}. The theory is a potential
candidate of quantum field theory of gravity. It assumes a
Lifshitz-like anisotropic scaling between space and time at short
distances, characterized by a dynamical critical exponent z = 3 and
thus breaking the Lorentz invariance. While in the IR limit it flows
to z = 1, retrieving the Einstein's General Relativity(GR). The
discussions on the consequences of theory are going
on\cite{padilla,sot}. As a new theory, it is interesting to
investigate its various aspects in parallel. Since the theory has
the same Newtonian and post Newtonian corrections as those of GR,
systems of strong gravity, like black holes, are needed to get
observable deviation from the standard GR. Various black hole
solutions are found in HL
theory\cite{lmp,park,bhsol1,bhsol2,bhsol3,bhsol4,bhsol5,bhsol6,bhsol7,bhsol8,bhsol9,bhsol10,bhsol11,bhsol12,bhsol13,bhsol14,bhsol15,bhsol16,bhsol17,bhsol18}
 of which one with asymptotically flat Minkowski spacetime is the KS black hole obtained by
applying deformation in the original theory\cite{ks}. Various
aspects of KS black hole were explored in the
past\cite{ksasp1,ksasp2,ksasp3,ksasp4,ksasp5,ksasp6,ksasp7,ksasp8,ksasp9,ksasp10,ksasp11,ksasp12,ksasp13}.

Perturbation of black hole is the best way to study the properties
of these exotic objects. As it is well established, the evolution of
perturbations in black hole space times involves three
stages\cite{kokko,nollert,wang1}. The first one is an initial
response containing the information of the initial form of the
original wave field followed by a region dominated by damped
oscillation of the field called quasinormal modes, which depends
entirely on the background black hole spacetimes. Finally the field
decays at late time as a power-law fall off of the field. QNMs play
an important role in astrophysical processes involving black holes.
These interests are associated with their relevance in gravitational
wave analysis\cite{ferrari}. Apart from the observational interest
of detection of quasinormal ringing by gravitational wave
detectors\cite{berti}, study of QNM of AdS black holes are of
particular interests in the study of AdS/CFT
correspondence\cite{malda1,malda2}. It was conjectured that a large
static black hole in AdS space corresponds to a thermal state in
conformal field theory (CFT). The fundamental link between QNMs of a
large AdS black hole and the dual field theory were shown in the
work of Horowitz who computed the QNMs of AdS black
holes\cite{horowitz}. Motivated by this, a series of studies on the
QNMs in asymptotically AdS backgrounds were done for various fields
and dimensions\cite{wang2,cardoso1,wang3,cardoso2}. See
Ref.\cite{berti2} for a recent review of QNMs of the black hole. The
late-time relaxation of field perturbation is also an interesting
topic of study since it reveals the actual physical mechanism by
which a perturbed black hole sheds its hairs. It was first
demonstrated by Price that in the background of the Schwarzschild
spacetime the massless neutral field at late time dies off as
$\Psi\sim t^{-(2\ell+2)}$ or $\Psi\sim t^{-(2\ell+3)}$ depending on
the initial conditions and the multipole order $\ell$ \cite{price1}.
The evolution of field and the spectra of QNMs may be different in
various theories of gravity and would help us to distinguish these
theories.

Dynamical evolution of scalar field in Horava black hole spacetimes
is studied using Horowitz-Hubeny approach\cite{ding}. The spectrum
of entropy/area is discussed from the viewpoint of QNMs of scalar
field in HL gravity\cite{majhi,setare}. QNMs of various fields
around KS black hole spacetime were calculated using WKB
method\cite{songbai,konoplyaqnm,wang,lin}. In earlier work\cite{nv}
we have studied time evolution of scalar field using the time domain
integration method\cite{gundlach,wang3,zhidenko}. A noticeable
deviation of the evolution behavior of scalar field in the ringdown
region from the standard Schwarzschild black hole was found. However
the late time evolution of field fails to show any distinction from
the Schwarzschild case. The late time behavior of massless scalar
field depends only on the multipole order $\ell$ and decays as
$t^{-(2\ell+3)}$. Where as massive field in the intermediate range
the decays as $t^{-(\ell+3/2)}sin(mt)$, but in the asymptotic late
time the decay is dominated by $t^{-5/6}sin(mt)$ tail.

It is interesting to see how various other field perturbations decay
in KS black hole spacetime. In this paper we study the evolution of
electromagnetic and massless Dirac fields in the KS black hole
spacetime and probe the signature of the now theory by comparing the
present results with those of the standard theory of GR. The paper
is organized as follows. In Sec.\ref{sec2} we derive the radial wave
equation for various field perturbations around KS black hole and
discuss the numerical method used. In Sec.\ref{sec3} the evolution
of electromagnetic and massless Dirac field are studied using the
method of time domain integration and the QNMs are evaluated and the
results are discussed in Sec.\ref{sec4}.

\section{Perturbation of fields around KS black hole}
\label{sec2} The IR vacuum of pure HL gravity is found to be anti-de
Sitter\cite{lmp,park}. Even though HL gravity could recover GR in IR
at the action level for a particular value of the parameter
$\lambda=1$, there found a significant difference between these
black hole solutions and the usual Schwarzschild AdS solution. The
asymptotic fall-off of the metric for these black hole solutions is
much slower than that of usual Schwarzschild AdS black holes in GR.
Meanwhile Kehagias and Sfetsos\cite{ks} could find a black hole
solution in asymptotically flat Minkowski spacetimes by applying
deformation in HL theory by adding a term proportional to the Ricci
scalar of three-geometry, $\mu^{4}R^{(3)}$ while the cosmological
constant $\Lambda_{W}\rightarrow0$.  This will not alter the UV
properties of the theory but it does the IR ones leading to
Minkowski vacuum analogous to Schwarzschild spacetime in GR. KS
solution is given as\cite{ks},
\begin{equation}
\ ds^{2}=-N(r)^{2}dt^{2}+f(r)^{-1}dr^{2}+r^{2}d\Omega ^{2},
\label{eq1}
\end{equation}

where $f(r)$ has the form,
\begin{equation}
N(r)^{2}=f(r)=\frac{2(r^{2}-2Mr+\alpha)}{r^{2}+2\alpha+\sqrt{r^{4}+8\alpha
Mr}}. \label{fofr}
\end{equation}

The event horizons are at,
\begin{equation}
r_{\pm}=M\pm \sqrt{M^{2}-\alpha}. \label{hori}
\end{equation}
When $\alpha=0$ the solution reduces to the Schwarzschild spacetime
case.

The evolution of massless scalar field $\Phi$, electromagnetic field
$A_{\mu}$ and massless Dirac field $\psi$ in the spacetime
$g_{\mu\nu}$, specified by Eq.(\ref{eq1}) are governed by the
Klein-Gordon\cite{nv}, Maxwell's\cite{ruffini} and the Dirac
equations\cite{birrell} respectively,

\begin{equation}
\frac{1}{\sqrt{-g}}\partial_{\mu}(\sqrt{-g}g^{\mu\nu}\partial_{\nu})\Phi=0,
\label{KGeqn}
\end{equation}

\begin{equation}
F_{;\nu}^{\mu\nu}=0,\qquad with\qquad
F_{\mu\nu}=A_{\nu;\mu}-A_{\mu;\nu}, \label{Maxeqn}
\end{equation}

\begin{equation}
\left[ \gamma ^{a}e_{a}^{\mu }(\partial _{\mu }+\Gamma _{\mu
})\right] \psi =0.  \label{Deqn}
\end{equation}

The radial part of the above perturbation equations can be decoupled
from their angular parts and reduced to form,

\begin{equation}
\left(-\frac{\partial^{2}}{\partial
t^{2}}+\frac{\partial^{2}}{\partial
r^{2}_{*}}\right)\Psi_{\ell}(t,r)=-V(r)\Psi_{\ell}(t,r)=0,
\label{waveqn}
\end{equation}
where $r_{*}$ is the tortoise coordinate defined by
$dr_{*}=\frac{1}{f}dr$ and the effective potentials $V(r)$ for
different fields are given by,

\begin{equation}
V_{S}(r)=f(r)\left [
\frac{\ell(\ell+1)}{r^{2}}+\frac{1}{r}\frac{\partial f(r)}{\partial
r}\right ], \qquad \ell=0,1,2,..., \label{scpot}
\end{equation}

\begin{equation}
V_{EM}(r)=f(r)\left [ \frac{\ell(\ell+1)}{r^{2}}\right ], \qquad
\ell=1,2,3,..., \label{empot}
\end{equation}

\begin{equation}
V_{D\pm}(r)=\frac{\sqrt{f(r)}|k|}{r^{2}}\left[
|k|\sqrt{f(r)}\pm\frac{r}{2}\frac{\partial f(r)}{\partial r}\mp
f(r)\right],\qquad |k|=1,2,3,...,  \label{Dpot}
\end{equation}

Here $k$ is positive or negative nonzero integer related to the
total orbital angular momentum by $\ell=|k+\frac{1}{2}|-\frac{1}{%
2}$ and $V_{D+}$ and $V_{D-}$ are the super symmetric partners and
give same spectra. So we choose $V_{D+}$ by omitting the subscript.
The perturbation equations (Eq.(\ref{waveqn})) can be written in the
null coordinates $u=t-r^{*}$ and $v=t+r^{*}$ as,
\begin{equation}
-4\frac{\partial^{2}}{\partial u \partial
v}\Psi(u,v)=V(u,v)\Psi(u,v). \label{waveqn2}
\end{equation}
In order to get the time evolution picture of the field we integrate
this equation numerically using the following finite difference
scheme suggested in\cite{gundlach},

\begin{equation}
\Psi_{N}=\Psi_{W}+\Psi_{E}-\Psi_{S}-\frac{h^{2}}{8}V(S)(\Psi_{W}+\Psi_{E})+O(h^{4})
\end{equation}

where the points N, S, E and W form a null rectangle with an overall
grid scale factor of $h$ having relative positions as $N(u+h,v+h)$,
$W(u+h,v)$, $E(u,v+h)$ and $S(u,v)$. Since we are interested in the
late-time behavior of the wave function which is found to be
independent of the initial shape, we set $\psi(u,v=0)=0$ and a
Gaussian profile $\Psi(u=0,v)=exp\left
[-\frac{(v-v_{c})^2}{2\sigma^2} \right ]$. In all our calculations
we set the initial Gaussian with width $\sigma=3$ centered at
$v_{c}=10$. In order to progress on the above numerical scheme one
has to get the value of the potential at $r(r_{*})=r((v-u)/2)$ at
each step. Here we use the Runge-Kutta method to numerically
integrate the equation for the tortoise coordinate and find the
values of $r(r_{*})$ at each step by cubic spline interpolation as
suggested in Ref.\cite{zhidenko}. After the integration is completed
the values of $\Psi$ on line $u=v-2c$ is extracted and plotted as a
function of $t=v-c$; here, c is the value of $r_{*}$ at which the
field is evaluated.

\section{Time evolution of fields around KS black hole}
\label{sec3} 
\begin{figure}[h]
\centering
\includegraphics[width=0.6\columnwidth]{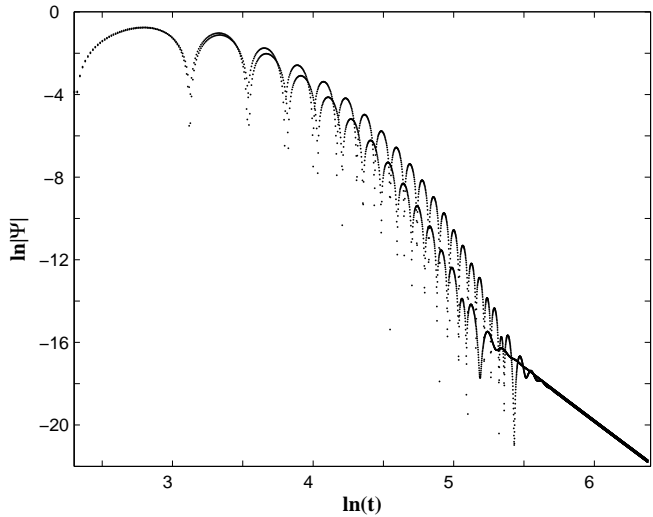}%
\caption{Time evolution of massless scalar field around KS black
hole with $\alpha=0.8$(top curve) and the Schwarzschild black
hole(bottom curve) for $\ell=1$.
} %
\label{sfig1}
\end{figure}

The evolution of scalar field around KS black hole was
studied in Ref. \cite{nv}. The typical nature of time evolution of a
massless scalar field is shown in Fig.\ref{sfig1}. The plot shows
how the generic behavior of wave function (initial outburst,
quasi-normal oscillations, and power-law decay) in HL theory is
differed from pure Schwarzschild spacetime. The QNM phase ends at a
later time in HL theory and the oscillation frequency and the
damping time have a higher values in HL theory. We can calculate the
QNM frequencies from the numerically integrated data by nonlinear
$\chi^{2}$ fitting. The calculated values of QNMs are given in
Table\ref{ta1} and Table\ref{ta2}.

\begin{table}[h]
\begin{center}
\begin{tabular}{p{0.4in}p{0.7in}p{0.7in}p{0.7in}p{0.7in}}
\hline\hline
& \multicolumn{2}{c}{WKB} & \multicolumn{2}{|c}{Time domain} \\
$\alpha$ & $Re(\omega)$ & $Im(\omega)$ & \multicolumn{1}{|r}{$Re(\omega)$} & $Im(\omega)$ \\
\hline
0 & 0.29111 & -0.09800 & 0.29224 & -0.09758 \\
0.2 & 0.29564 & -0.09424 & 0.29750 & -0.09418 \\
0.4 & 0.30071 & -0.09008 & 0.30154 & -0.09027 \\
0.5 & 0.30345 & -0.08777 & 0.30474 & -0.08817 \\
0.6 & 0.30636 & -0.08526 & 0.30850 & -0.08564 \\
0.8 & 0.31260 & -0.07937 & 0.31416 & -0.07979 \\
1 & 0.31918 & -0.07162 & 0.32057 & -0.07247 \\
\hline\hline
\end{tabular}
\caption{QNM frequencies of massless scalar field for $\ell=1$,
calculated using WKB and numerical integration data} \end{center}
\label{ta1}
\end{table}

\begin{table}[h]
\begin{center}
\begin{tabular}{p{0.4in}p{0.7in}p{0.7in}p{0.7in}p{0.7in}}
\hline\hline
& \multicolumn{2}{c}{WKB} & \multicolumn{2}{|c}{Time domain} \\
$\alpha$ & $Re(\omega)$ & $Im(\omega)$ & \multicolumn{1}{|r}{$Re(\omega)$} & $Im(\omega)$ \\
\hline
0 & 0.48321 & -0.09681 & 0.48332 & -0.09676 \\
0.2 & 0.49085 & -0.09351 & 0.49087 & -0.09356 \\
0.4 & 0.49942 & -0.08969 & 0.49911 & -0.08981 \\
0.5 & 0.50412 & -0.08752 & 0.50266 & -0.08767 \\
0.6 & 0.50915 & -0.08511 & 0.50831 & -0.08527 \\
0.8 & 0.52039 & -0.07926 & 0.52029 & -0.07947 \\
1 & 0.53356 & -0.07108 & 0.53272 & -0.07138 \\
\hline\hline
\end{tabular}
\caption{QNM frequencies of massless scalar field for $\ell=2$,
calculated using WKB and numerical integration data}
\end{center}\label{ta2}
\end{table}

\begin{figure}[h]
\centering
\includegraphics[width=0.45\columnwidth]{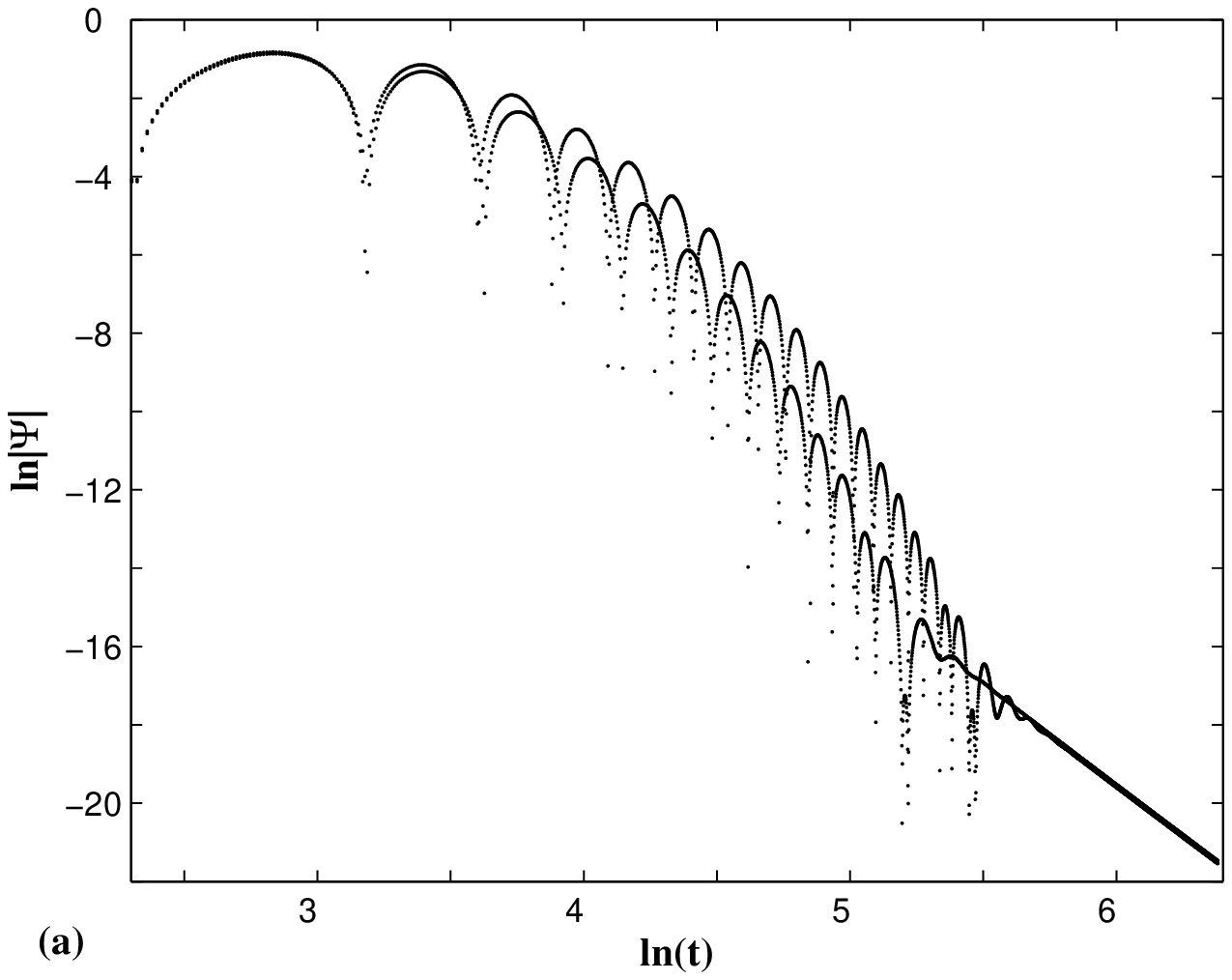}%
\hspace{0.15in}%
\includegraphics[width=0.45\columnwidth]{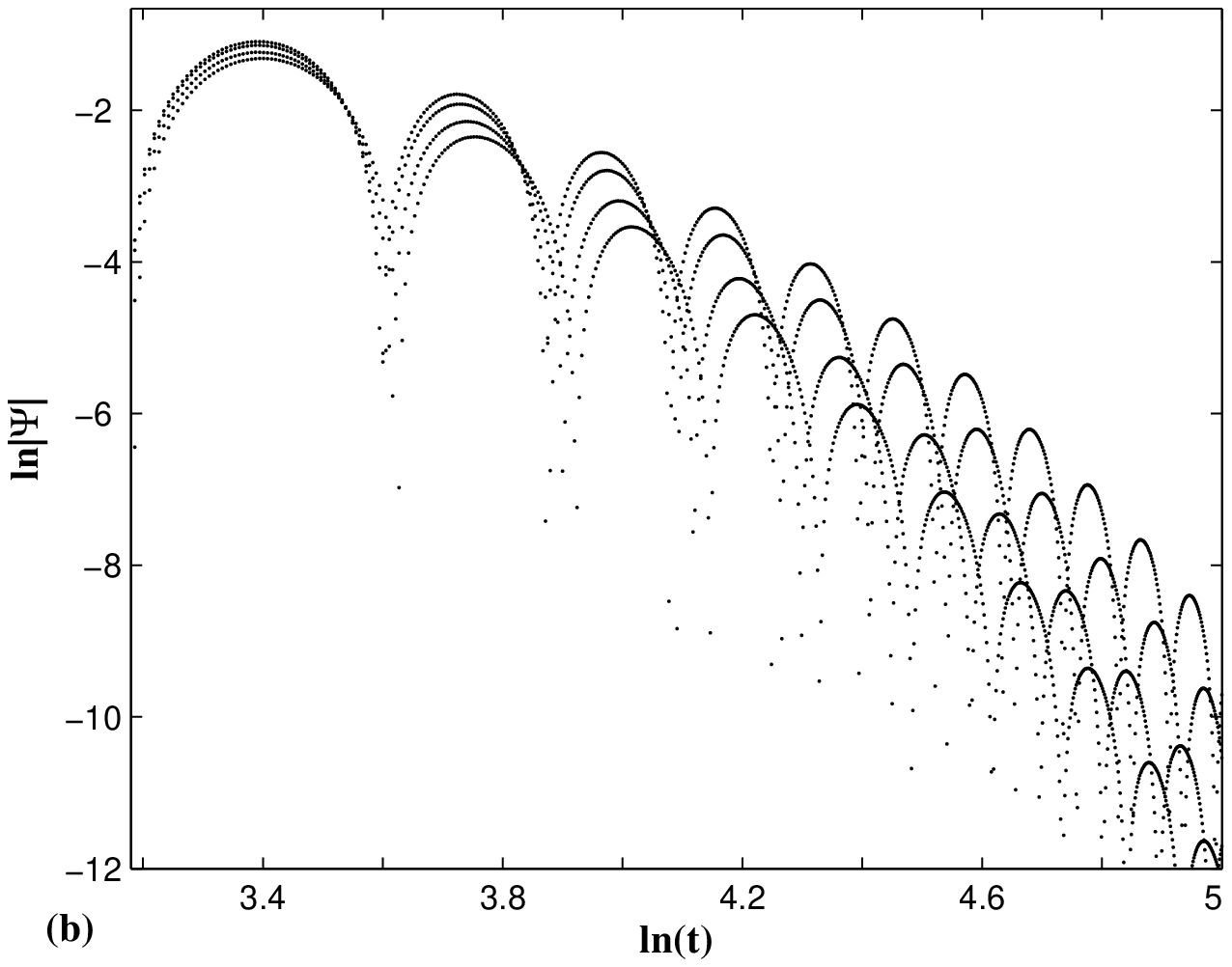}
\caption{Evolution of electromagnetic field for $\ell=1$. $(a)$
Around KS black hole with $\alpha=0.8$(top curve) and the
Schwarzschild black hole(bottom curve). $(b)$QNM region of the time
evolution for different values of $\alpha$. Curves from bottom to
top is for $\alpha = 0, 0.4, 0.8$ and $1$.
} %
\label{efig1}
\end{figure}

In order to study the evolution of electromagnetic field in HL black
hole spacetime we numerically integrate the perturbation equation
Eq.(\ref{waveqn2}) with effective potential Eq.(\ref{empot}) using
the time domain method. In Fig.\ref{efig1}(a) the time evolution of
the electromagnetic field $\Psi(t,r)$ at a fixed radius $r_{*}=10$
for $\ell=1$ is plotted in comparison with the corresponding
Schwarzschild case. We can see that the time length of the QNM phase
increases in HL theory. The late time tail starts at $t\approx300$
for KS black hole with $\alpha=0.8$ whereas it is at $t\approx230$
for the Schwarzschild case. The oscillation frequency and the
damping time have a higher values in HL theory. The variation of
oscillatory region of $\Psi(t,r)$ with the parameter $\alpha$ is
shown in Fig.\ref{efig1}(b). The oscillation frequency increases
with $\alpha$ and shows a slower damping for higher values of
$\alpha$.

The late-time behavior of the field around black hole spacetime is
also an important topic of study in black hole physics. It was found
that during the late-time a massless neutral field dies off as an
inverse power of time by the factor $t^{(2\ell+3)}$ depending on the
multipole order of the perturbation \cite{price1,price2}. It
interesting to see how the field at late time behaves in the HL
theory.

\begin{figure}[h]
\centering
\includegraphics[width=0.45\columnwidth]{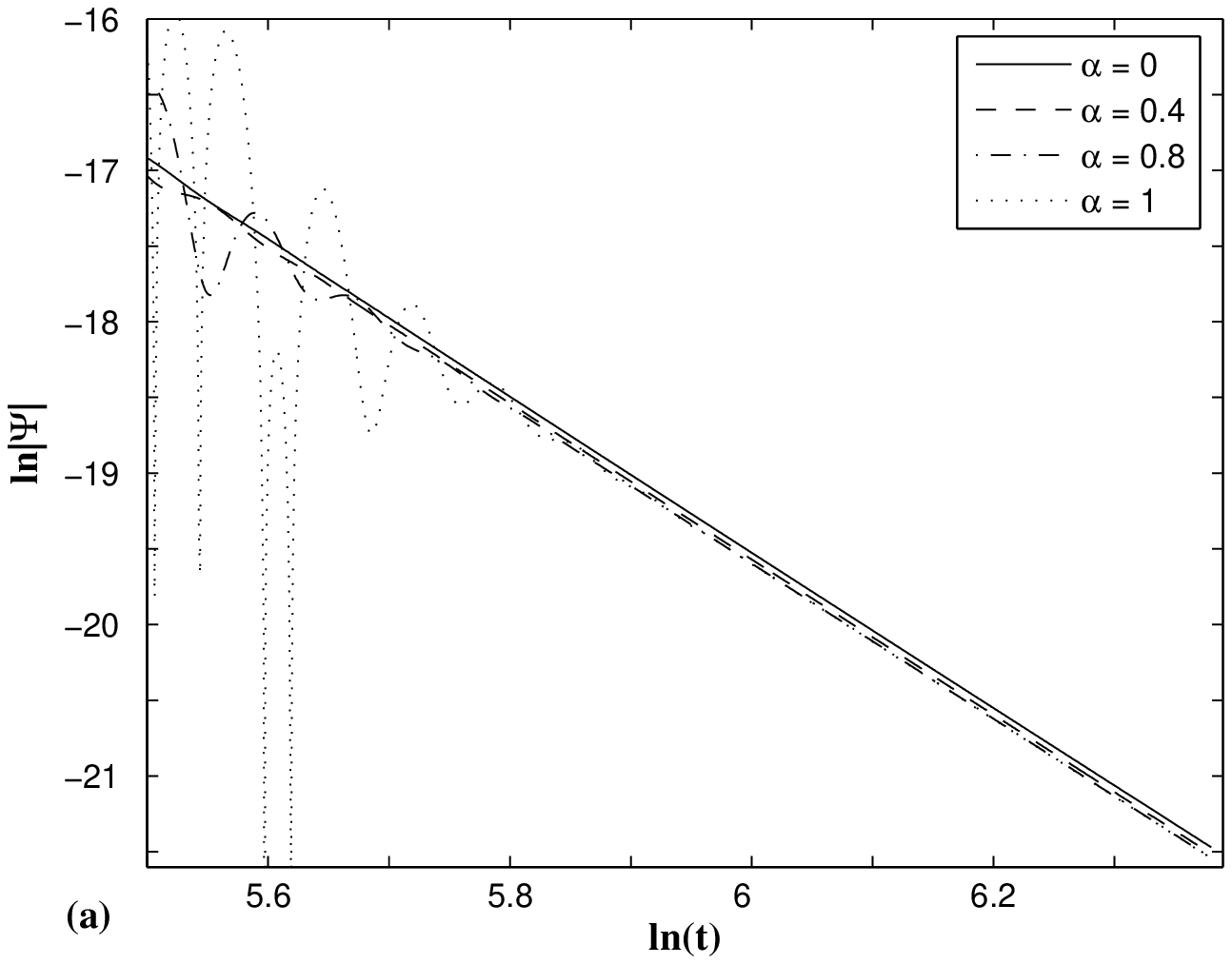}%
\hspace{0.15in}%
\includegraphics[width=0.45\columnwidth]{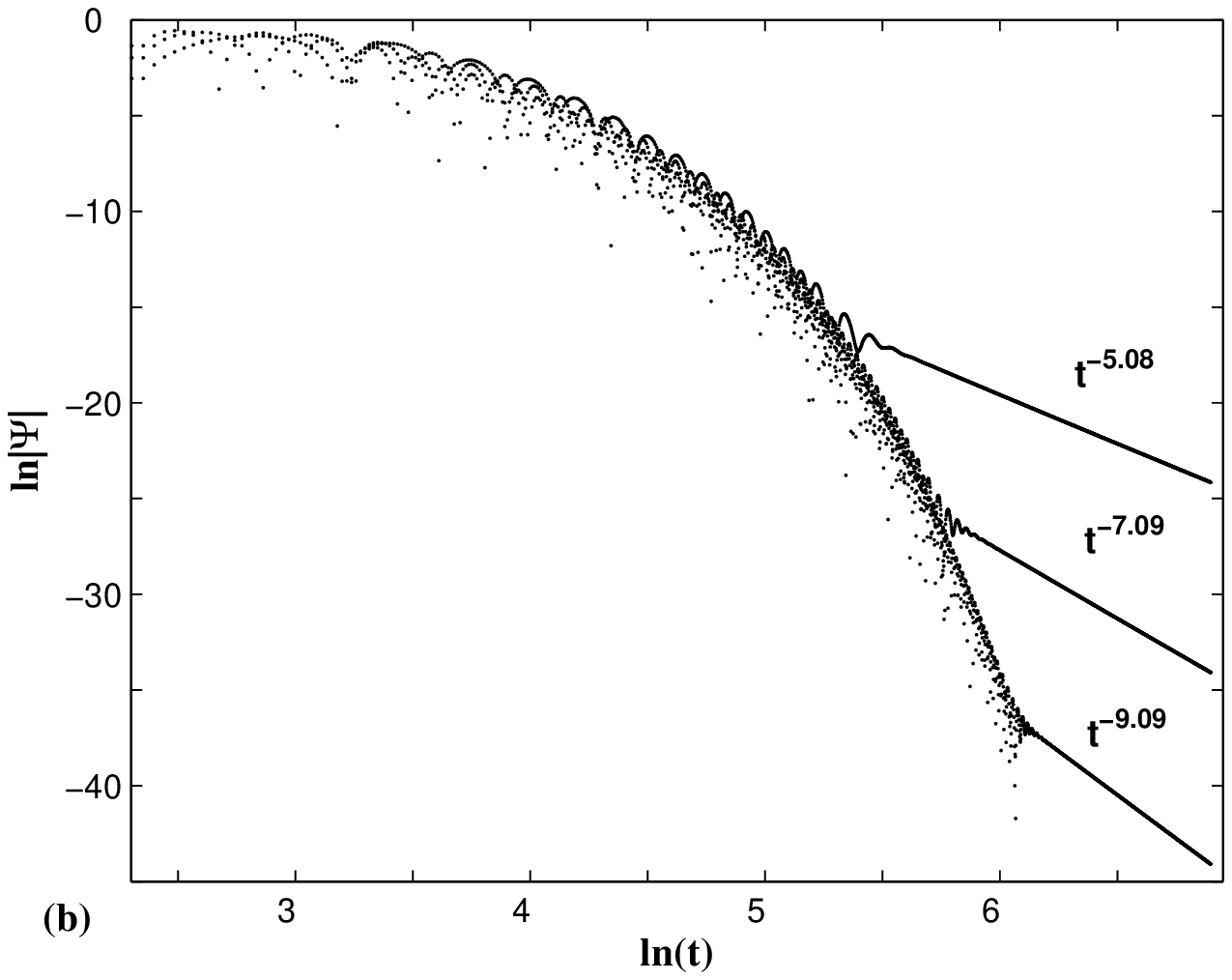}
\caption{Late time decay of electromagnetic field. $(a)$ for
different values of $\alpha$ with $\ell=1$. The Field decay as an
inverse power of time with $t^{-5.08}$, for all values of $\alpha$.
$(b)$Decay of field with different multipole order $\ell$ for
$\alpha=0.5$. The field decay as $t^{-5.08},t^{-7.09}$ and
$t^{-9.09}$ for $\ell=1,2$ and $3$.
} %
\label{efig2}
\end{figure}

Fig.\ref{efig2}(a) shows the late time behavior of wave function for
different values of $\alpha$, with multipole index $\ell=1$. We find
that the late time behavior is independent of $\alpha$ and follows
the behavior of the Schwarzschild case with $\Psi\sim t^{-5.1}$. In
Fig.\ref{efig2}(b) field evolution for different multipole index is
shown with $\alpha=0.5$. The perturbation dies off at late time as
$t^{-(2\ell+3)}$ as in the case of Schwarzschild case. The field
falls off as $\Psi\sim t^{-5.08},t^{-7.09}$ and $t^{-9.09}$ for
$\ell=1,2$ and $3$ respectively. The predicted values are -5, -7 and
-9 respectively.

We have seen that quasinormal ringing phase is a dominated form of
decay in the evolution of perturbations after the initial transient
phase. The QNMs calculated from the numerically integrated data are
given in Table\ref{ta3} and Table\ref{ta4}. We can find a good
agreement with the earlier results obtained using WKB method in
Ref.\cite{lin}. Results show that the oscillation frequency and the
damping time increase with $\alpha$.

\begin{table}[h]
\begin{center}
\begin{tabular}{p{0.4in}p{0.7in}p{0.7in}p{0.7in}p{0.7in}}
\hline\hline
& \multicolumn{2}{c}{WKB} & \multicolumn{2}{|c}{Time domain} \\
$\alpha$ & $Re(\omega)$ & $Im(\omega)$ & \multicolumn{1}{|r}{$Re(\omega)$} & $Im(\omega)$ \\
\hline
0 & 0.24587 & -0.09311 & 0.24401 & -0.09051 \\
0.2 & 0.25152 & -0.08942 & 0.25225 & -0.08706 \\
0.4 & 0.25796 & -0.08530 & 0.25733 & -0.08447 \\
0.5 & 0.26151 & -0.08299 & 0.26383 & -0.08135 \\
0.6 & 0.26532 & -0.08047 & 0.26575 & -0.08009 \\
0.8 & 0.27375 & -0.07435 & 0.27771 & -0.07454 \\
1 & 0.28308 & -0.06569 & 0.28478 & -0.06581 \\
\hline\hline
\end{tabular}
\caption{QNM frequencies of electromagnetic field for $\ell=1$,
calculated using WKB and numerical integration data}
\end{center}\label{ta3}
\end{table}

\begin{table}[h]
\begin{center}
\begin{tabular}{p{0.4in}p{0.7in}p{0.7in}p{0.7in}p{0.7in}}
\hline\hline
& \multicolumn{2}{c}{WKB} & \multicolumn{2}{|c}{Time domain} \\
$\alpha$ & $Re(\omega)$ & $Im(\omega)$ & \multicolumn{1}{|r}{$Re(\omega)$} & $Im(\omega)$ \\
\hline
0 & 0.45713 & -0.09506 & 0.45088 & -0.09471 \\
0.2 & 0.46531 & -0.09179 & 0.46239 & -0.09117 \\
0.4 & 0.47453 & -0.08799 & 0.47153 & -0.08711 \\
0.5 & 0.47961 & -0.08582 & 0.47835 & -0.08502 \\
0.6 & 0.48507 & -0.08339 & 0.48332 & -0.08183 \\
0.8 & 0.49736 & -0.07743 & 0.49756 & -0.07439 \\
1 & 0.51197 & -0.06887 & 0.52823 &  -0.06897 \\
\hline\hline
\end{tabular}
\caption{QNM frequencies of electromagnetic field for $\ell=2$,
calculated using WKB and numerical integration data}
\end{center}\label{ta4}
\end{table}

\begin{figure}[h]
\centering
\includegraphics[width=0.45\columnwidth]{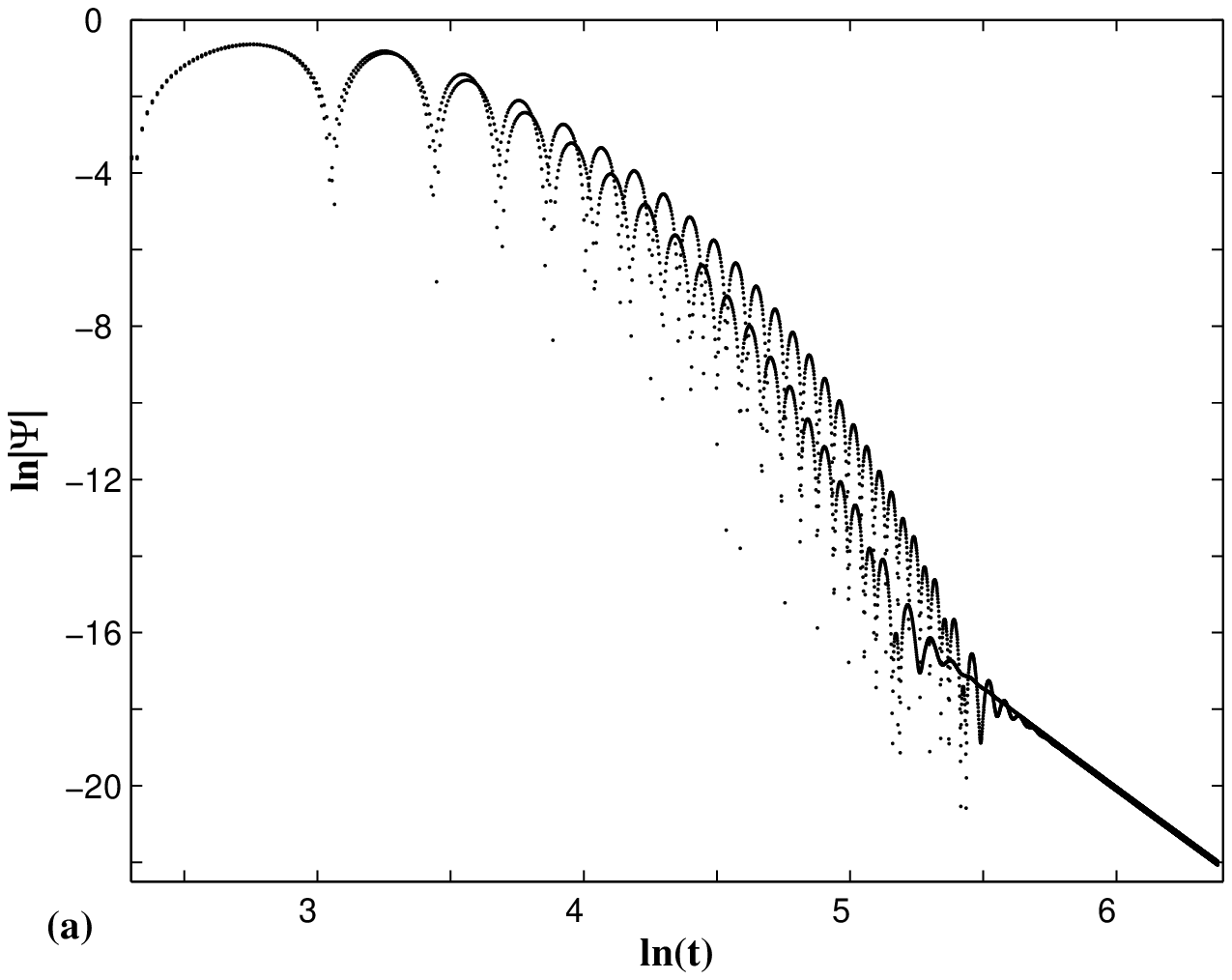}%
\hspace{0.15in}%
\includegraphics[width=0.45\columnwidth]{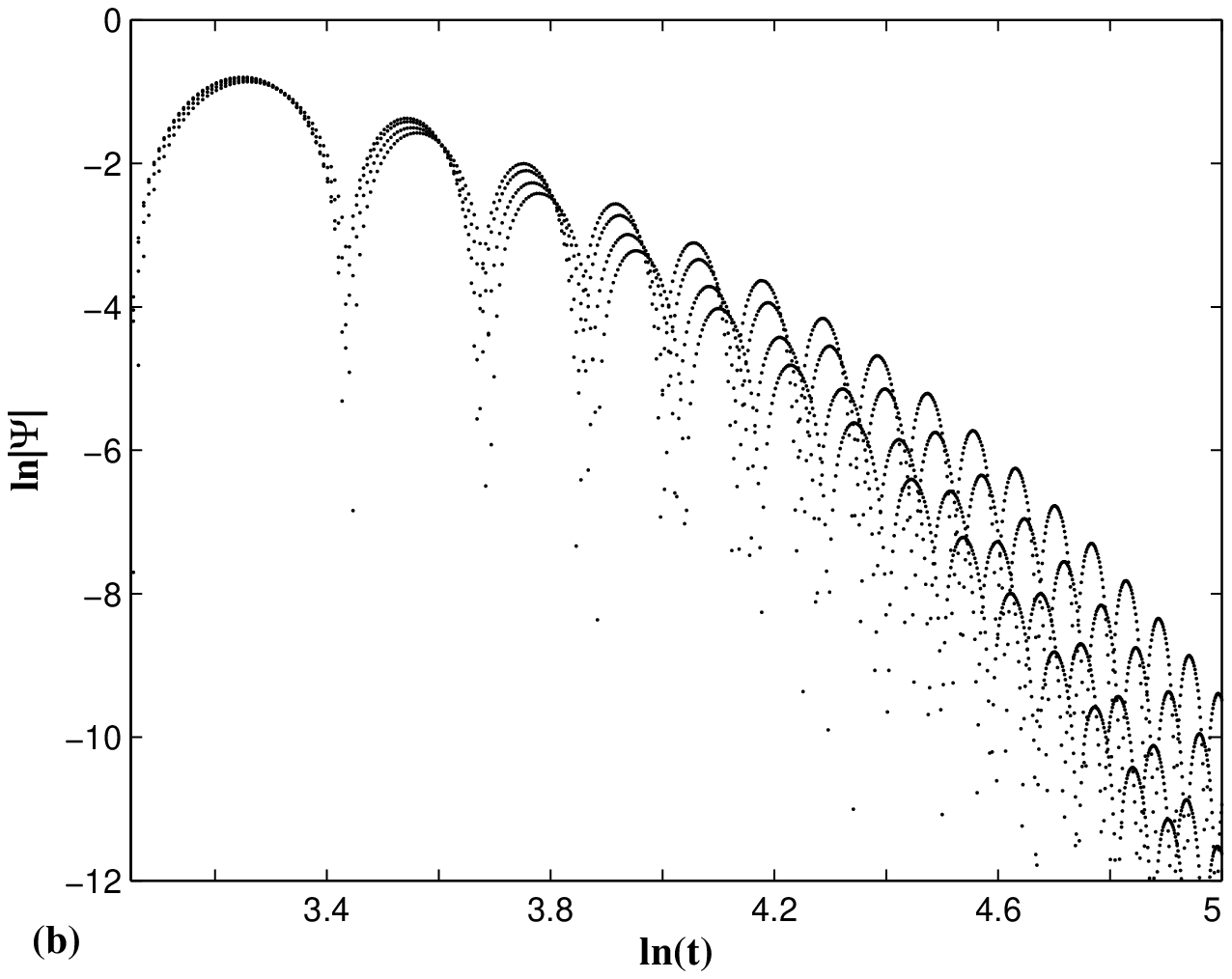}
\caption{Time evolution of Dirac field with $k=2$. $(a)$ Around KS
black hole(top curve) and the Schwarzschild black hole(bottom
curve). $(b)$QNM region of the time evolution of Dirac field for
different values of $\alpha$. Curves from bottom to top is for
$\alpha = 0, 0.4, 0.8$ and $1$.
} %
\label{dfig1}
\end{figure}

Now we study the evolution of Dirac field in HL black hole
spacetime. We numerically integrate the perturbation equation
Eq.(\ref{waveqn2}) with the effective potential Eq.(\ref{Dpot})
using the time domain method. Fig.\ref{dfig1}(a) displays the time
evolution of the Dirac wave function $\Psi(t,r)$ at a fixed radius
$r_{*}=10$, for $k=2$. The deviation of generic time dependence of
wave function in the HL theory from pure Schwarzschild spacetime is
clear in the plot. QNM region lasts for a longer time in HL theory.
The late time tail starts at $t\approx310$ for $\alpha=0.8$ whereas
it is at $t\approx240$ for the Schwarzschild case. Also the
oscillation frequency and the damping time have a higher values in
HL theory. The variation of oscillatory region of $\Psi(t,r)$ with
the parameter $\alpha$ is shown in Fig.\ref{dfig1}(b). The
oscillation frequency increases with $\alpha$ and shows a slower
damping for higher values of $\alpha$.

\begin{figure}[h]
\centering
\includegraphics[width=0.45\columnwidth]{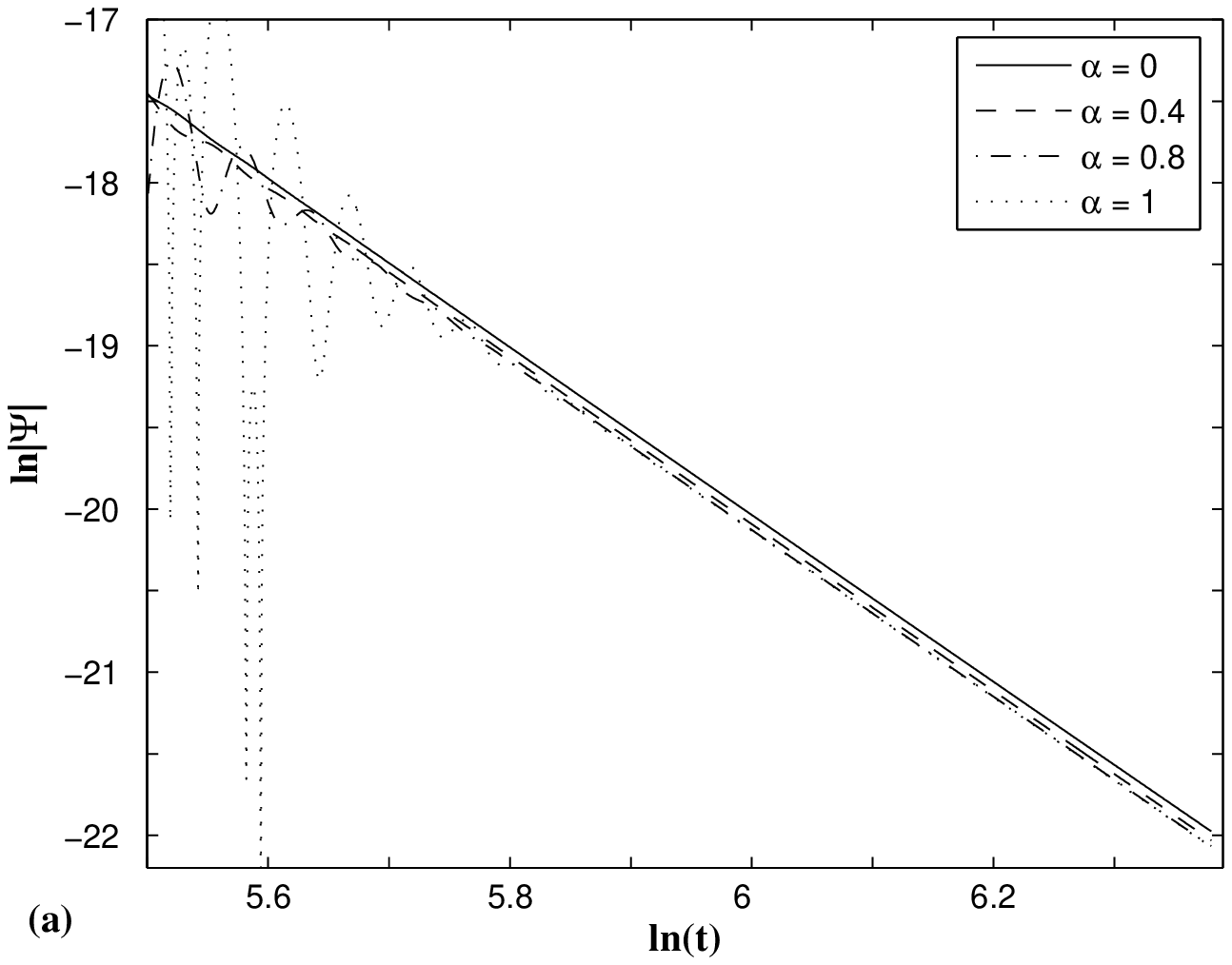}%
\hspace{0.15in}%
\includegraphics[width=0.45\columnwidth]{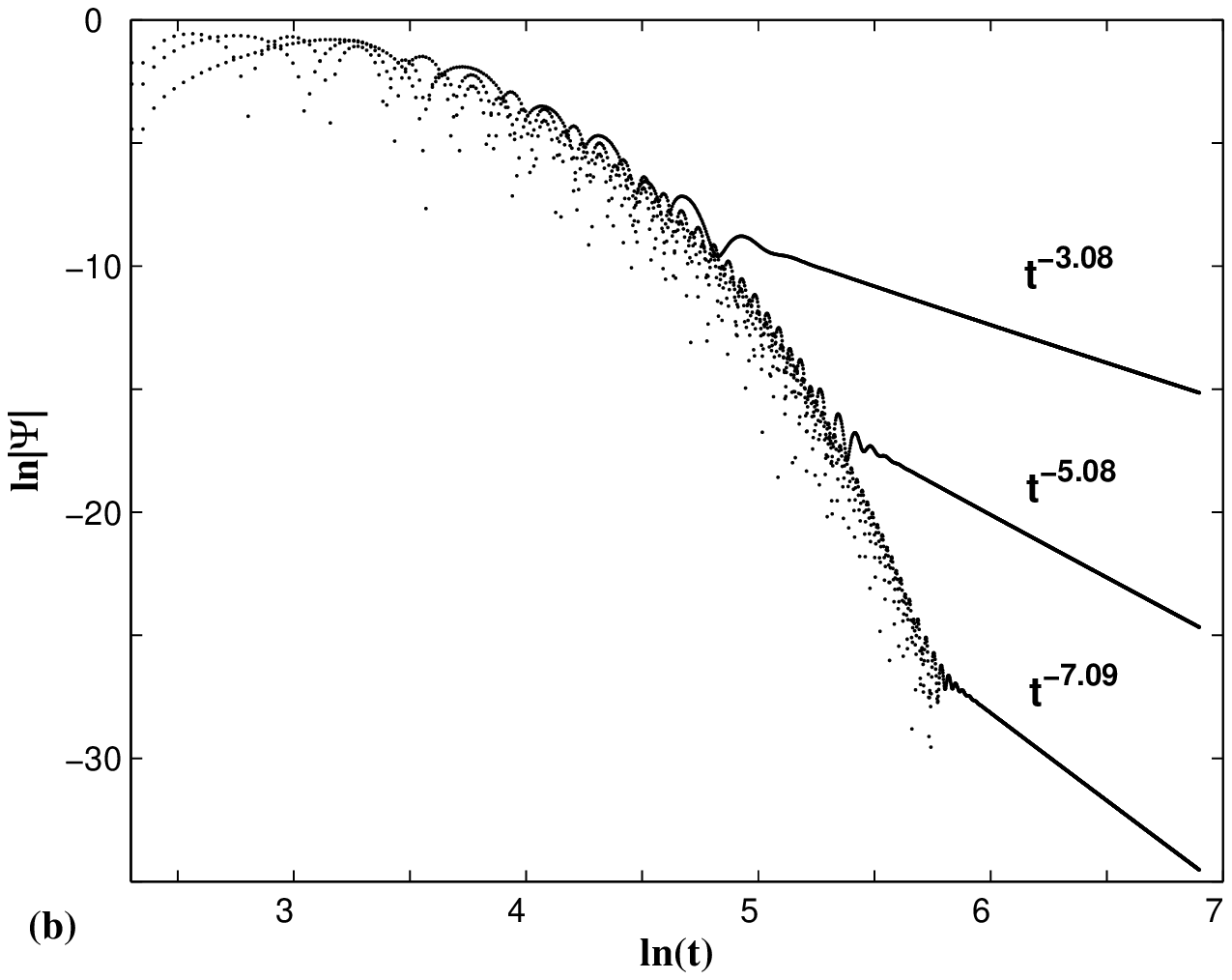}
\caption{Late time decay of Dirac field. $(a)$ For different values
of $\alpha$ with $k=2$. The field decay as an inverse power law of
time with $t^{-5.08}$ for all values of $\alpha$. $(b)$Decay of
Dirac field for different $k$ with $\alpha=0.5$. The field decay as
$t^{-3.08},t^{-5.08}$ and $t^{-7.09}$ for $k=1,2$ and $3$
respectively.
} %
\label{dfig2}
\end{figure}

Fig.\ref{dfig2}(a) shows the late time behavior wave function for
different values of $\alpha$, with $k=2$. The late time behavior is
independent of $\alpha$ and field decays in the inverse power of
time as $t^{-5.08}$. In Fig.\ref{dfig2}(b), field evolution for
different multipole indices are shown with $\alpha=0.5$. The
perturbation dies off at late time as $\Psi\sim t^{-3.08},t^{-5.08}$
and $t^{-7.09}$ for $k=1,2$ and $3$ respectively. The predicted
values are -3, -5 and -7 respectively.

The quasinormal ringing phase of the perturbation can be seen
clearly in these figures. The calculated values of QNMs from the
time domain data are given in Table\ref{ta5} and Table\ref{ta6}. We
can find a good agreement of the values obtained in Ref.\cite{wang}
from the WKB scheme. Results show that the oscillation frequency and
the damping time increase with $\alpha$.

\begin{table}[h]
\begin{center}
\begin{tabular}{p{0.4in}p{0.7in}p{0.7in}p{0.7in}p{0.7in}}
\hline\hline
& \multicolumn{2}{c}{WKB} & \multicolumn{2}{|c}{Time domain} \\
$\alpha$ & $Re(\omega)$ & $Im(\omega)$ & \multicolumn{1}{|r}{$Re(\omega)$} & $Im(\omega)$ \\
\hline
0 & 0.17645 & -0.10011 & 0.17830 & -0.10744 \\
0.2 & 0.18187 & -0.09614 & 0.17952 & -0.09481 \\
0.4 & 0.18709 & -0.09167 & 0.18756 & -0.09312 \\
0.5 & 0.18963 & -0.08916 & 0.19040 & -0.09249 \\
0.6 & 0.19212 & -0.08639 & 0.19333 & -0.09172 \\
0.8 & 0.19681 & -0.07981 & 0.19635 & -0.07572 \\
1 & 0.20053 & -0.07113 & 0.200101 & -0.06939 \\
\hline\hline
\end{tabular}
\caption{QNM frequencies of Dirac field for $k=2$, calculated using
WKB and numerical integration data}
\end{center}\label{ta5}
\end{table}

\begin{table}[h]
\begin{center}
\begin{tabular}{p{0.4in}p{0.7in}p{0.7in}p{0.7in}p{0.7in}}
\hline\hline
& \multicolumn{2}{c}{WKB} & \multicolumn{2}{|c}{Time domain} \\
$\alpha$ & $Re(\omega)$ & $Im(\omega)$ & \multicolumn{1}{|r}{$Re(\omega)$} & $Im(\omega)$ \\
\hline
0 & 0.37863 & -0.09654 & 0.37348 & -0.08850 \\
0.2 & 0.38489 & -0.09317 & 0.38126 & -0.09488 \\
0.4 & 0.39195 & -0.08932 & 0.38935 & -0.09014 \\
0.5 & 0.39583 & -0.08712 & 0.39126 & -0.08816 \\
0.6 & 0.39998 & -0.08468 & 0.39153 & -0.08388 \\
0.8 & 0.40920 & -0.07869 & 0.40337 & -0.07758 \\
1 & 0.41978 & -0.07021 & 0.41415 & -0.07239 \\
\hline\hline
\end{tabular}
\caption{QNM frequencies of Dirac field for $k=2$, calculated using
WKB and numerical integration data}
\end{center}\label{ta6}
\end{table}

\begin{figure}[h]
\centering
\includegraphics[width=0.48\columnwidth]{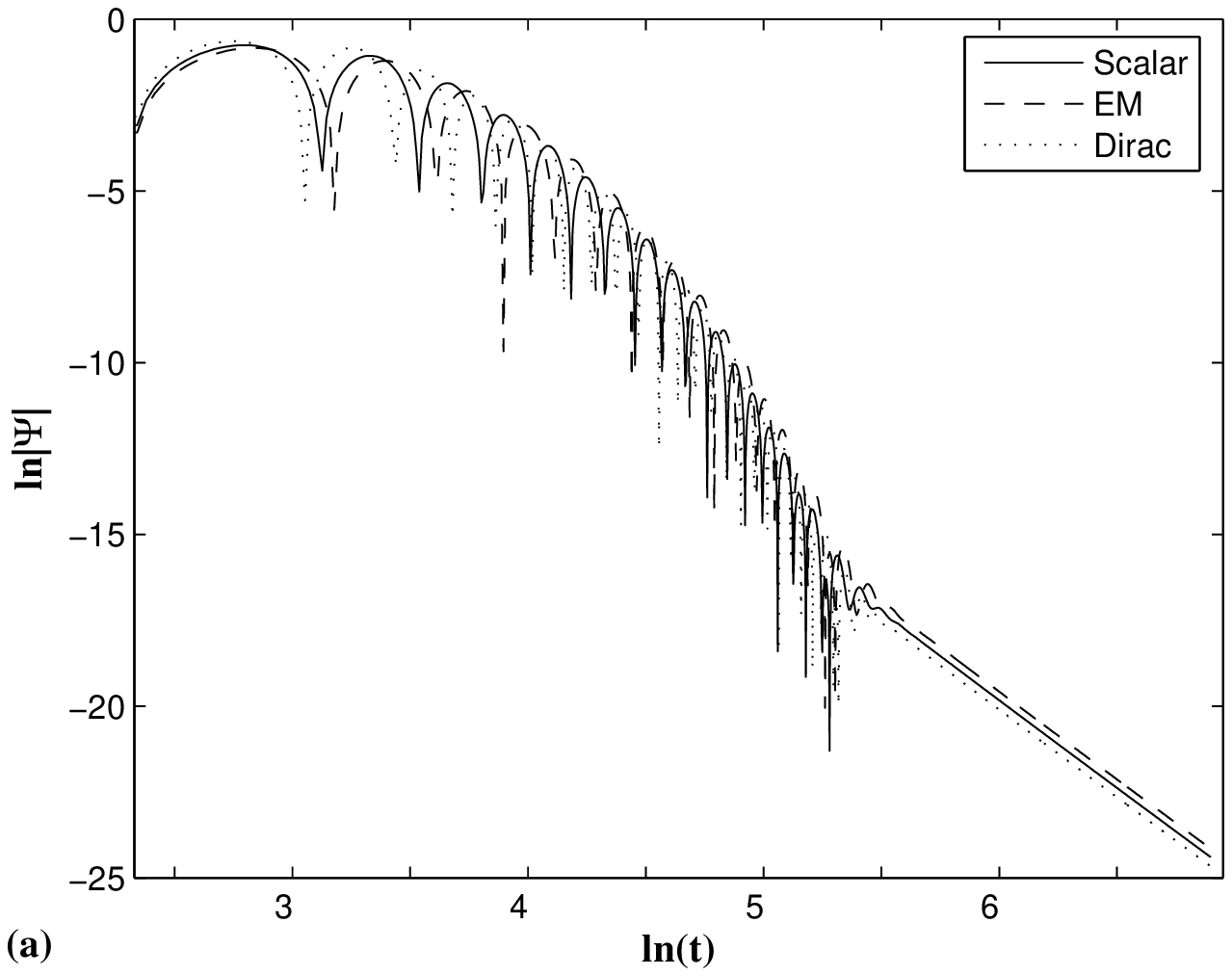}%
\hspace{0.1in}%
\includegraphics[width=0.48\columnwidth]{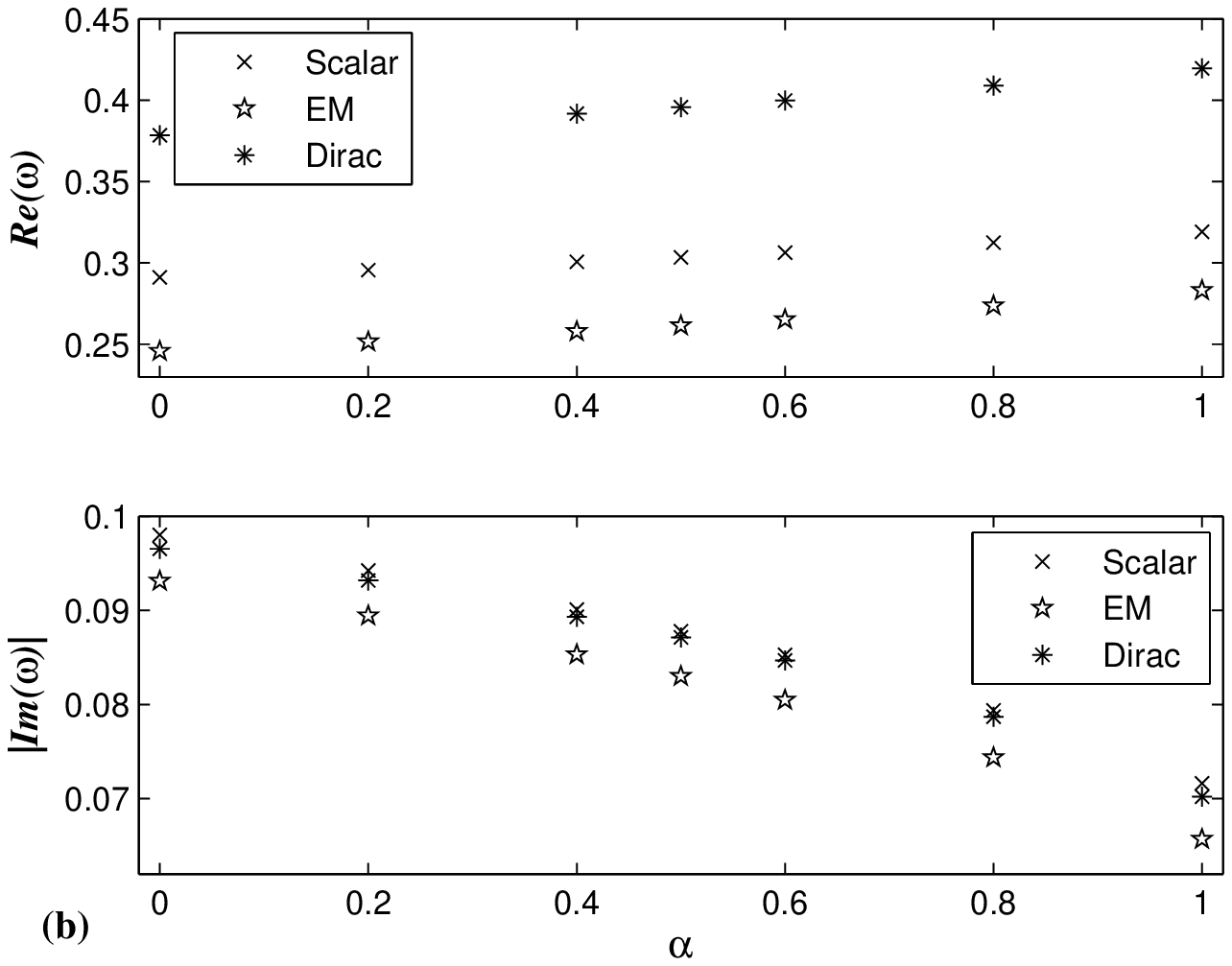}
\caption{$(a)$Evolution of massless scalar, Dirac and
electromagnetic fields around KS black hole with $\alpha=0.5$ and
$\ell=1$. $(b)$ QNMs of different field as a function of $\alpha$
with $\ell=1$.
} %
\label{sed}
\end{figure}

Finally we compare the time evolution of different fields in HL
gravity. In Fig.\ref{sed}(a) massless scalar, Dirac and
electromagnetic fields are plotted for $\alpha=0.5$. The only
difference between evolution of these three fields is in the QNM
phase whereas the late time tails follow the same decaying pattern
with same power law exponent. In Fig.\ref{sed}(b) the variation of
real and imaginary part of QNM versus $\alpha$ are plotted. All the
three fields show the same dependence on the HL parameter $\alpha$.

\section{Summary}
\label{sec4} The evolution of electromagnetic and massless Dirac
perturbations in KS black hole in the deformed  Ho\v{r}ava-Lifshitz
gravity is studied in this work. Time domain integration method is
used to obtain the evolution picture. The question raised was
whether we can distinguish the fundamental nature of the HL theory
from a knowledge of the evolution of fields around corresponding
black holes. Comparing with the Schwarzschild case, we have found
that the QNM phase has been extended for a longer time in HL theory
before the power-law tail begins. Also the oscillation frequency and
the damping time have higher values in HL theory. The variation of
the field evolution on the Horava parameter $\alpha$ are also
studied. In the time domain picture we find that the length of QNM
region increases with $\alpha$. The late time decay of field is
found to be independent of nature of field and the HL parameter
$\alpha$ and follows the same power-law tail behavior as in the case
of Schwarzschild black hole. These results will help us to check the
correctness of the theory provided our gravitational detectors can
trap these signals with high accuracy.

\section*{Acknowledgments}
NV wishes to thank UGC, New Delhi for financial support under RFSMS
scheme. VCK is thankful to CSIR, New Delhi for financial support
under Emeritus Scientistship scheme and wishes to acknowledge
Associateship of IUCAA, Pune, India. The authors are grateful to Dr.
Alexander Zhidenko for the help and clarifications received in
developing the code for the numerical integration.


\begin{thebibliography}{00}

\bibitem{horava1}P. Ho\v{r}ava, {\it Phys. Rev. D} {\bf79} 084008(2009).
\bibitem{horava2}P. Ho\v{r}ava, {\it JHEP} {\bf03} 020(2009).
\bibitem{horava3}P. Ho\v{r}ava, {\it Phys. Rev. Lett.} {\bf102} 161301(2009).
\bibitem{padilla}A Padilla, {\it J. Phys. Conf. Ser.} {\bf259} 012033(2010).
\bibitem{sot}T. P. Sotiriou, {\it J. Phys. Conf. Ser.} {\bf283} 012034(2011).

\bibitem{lmp} H. Lu, J Mei and C. N. Pope, {\it Phys. Rev. Lett.} \textbf{103} 091301(2009).
\bibitem{park}Mu-In Park, {it JHEP} \textbf{09} 123(2009).
\bibitem{bhsol1}H. Nastase, arXiv:0904.3604 [hep-th].
\bibitem{bhsol2} R. G. Cai, Y. Liu and Y. Sun , {\it JHEP} \textbf{06} 010(2009).
\bibitem{bhsol3} R. G. Cai, L. Cao and N. Ohta, {\it Phys. Rev. D} \textbf{80} 024003(2009).
\bibitem{bhsol4}A. Ghodsi and E. Hatefi, {\it Phys. Rev. D} \textbf{81} 044016(2010).
\bibitem{bhsol5}H. W. Lee Y. Kim and Y. S. Myung, arXiv:0907.3568
[hep-th].
\bibitem{bhsol6}J. Z. Tang and B. Chen, {\it Phys. Rev. D} \textbf{81} 043515
(2010).
\bibitem{bhsol7}E. Kiritsis and G. Kofinas, arXiv:0910.5487
[hep-th].
\bibitem{bhsol8}T. Harada, U. Miyamoto and N. Tsukamoto, {\it Int. J. Mod. Phys. D} {\bf 20} 111(2011).
\bibitem{bhsol9}E. Kiritsis, {\it Phys. Rev. D} \textbf{81} 044009(2010).
\bibitem{bhsol10}D. Capasso and A. Polychronakos, arXiv:0911.1535v3
[hep-th].
\bibitem{bhsol11}J. Tang, arXiv:0911.3849[hp-th].
\bibitem{bhsol12}H. W. Lee, Y. Kim, Y. S. Myung, {\it Eur. Phys. J. C} \textbf{70} 367(2010).
\bibitem{bhsol13}G. Koutsoumbas and P. Pasipoularides, arXiv:1006.3199v3 [hep-th].
\bibitem{bhsol14}J. Greenwald, A. Papazoglou and A. Wang, {\it Phys. Rev. D} \textbf{81} 084046(2010).
\bibitem{bhsol15}R. G. Cai and A. Wang, {\it Phys. Lett. B} \textbf{686} 166(2010).
\bibitem{bhsol16}G. Koutsoumbas, E. Papantonopoulos, P. Pasipoularides and M. Tsoukalas, arXiv:1004.2289v2 [hep-th].
\bibitem{bhsol17}A. N. Aliev and C. Senturk, {\it Phys. Rev. D.} {\bf 82} 104016(2010).
\bibitem{bhsol18}E. Gruss, arXiv:1005.1353v1 [hep-th].

\bibitem{ks}A. Kehagias and K. Sfetsos, {\it Phys. Lett. B} \textbf{678}(1) (2009) 123.

\bibitem{ksasp1}R. G. Cai, ,L. M. Cao and N. Ohta, {\it Phys. Lett.
B} \textbf{679} 504(2009).
\bibitem{ksasp2}Y. S. Myung, arXiv:0905.0957v2 [hep-th].
\bibitem{ksasp3}S. Chen, J. Jing, {\it Phys. Rev. D} \textbf{80}
024036(2009).
\bibitem{ksasp4}Y. S. Myuang, {\it Phys. Rev. D} \textbf{81} 064006(2010).
\bibitem{ksasp5}J. J. Peng and S. Q. Wu, {\it Eur. Phys. J. C} \textbf{66}
325(2010).
\bibitem{ksasp6}L. Iorio and M. L. Ruggiero, {\it Int. J. Mod. Phys. A} {\bf 25} 5399(2010).
\bibitem{ksasp7}L. Iorio and M. L. Ruggiero, {\it Open Astron. J.} {\bf 3}167(2010).
\bibitem{ksasp8}T. Harko, Z. Kovacs and F. S. N. Lobo, {\it Proc. R. Soc.} {\bf A467}
1390(2011).
\bibitem{ksasp9}T. Harko, Z. Kovacs and F. S. N. Lobo, {\it Phys. Rev. D} \textbf{80}
044021(2009).
\bibitem{ksasp10}M. R. Setare and M. Jamil, {\it Int. J. Theor. Phys.} {\bf 50} 511(2011).
\bibitem{ksasp11}W. Janke, D. A. Johnston and R. Kenna, arXiv:1005.3392v2 [hep-th].
\bibitem{ksasp12}B. Gwak and B. H. Lee, {\it JCAP} \textbf{09} 031(2010).
\bibitem{ksasp13}M. R. Setare and D. Momeni, {\it Int. J. Theor. Phys.} {\bf 49}
106(2011).

\bibitem{kokko}K. D. Kokkotas and B. G. Schmidt, {\it Living Rev. Relativ.} \textbf{2} 2(1999).
\bibitem{nollert}H. P. Nollert, {\it Class. Quantum Grav.} {\bf16} R159(1999)
\bibitem{wang1}B. Wang, {\it Braz. J. Phys.}, {\bf35},1029 (2005).
\bibitem{ferrari} V. Ferrari and L. Gualtieri {\it Gen. Rel. Grav.} \textbf{40} 945(2008).
\bibitem{berti}E. Berti, V. Cardoso and C. M. Will, {\it Phys. Rev. D}
{\bf73}064030(2006).
\bibitem{malda1}J. M. Maldacena, {\it Adv. Theor. Math. Phys.} {\bf2}
231(1998).
\bibitem{malda2}J. M. Maldacena, {\it Int. J. Theor. Phys.} {\bf 38} 1113(1999).

\bibitem{horowitz}G. T. Horowitz and V. E. Hubeny, {\it Phys. Rev. D}, {\bf62}, 024027(2000).
\bibitem{wang2}B. Wang, C. Y. Lin, and E. Abdalla, {\it Phys. Lett. B}, {\bf79}, 481(2000).
\bibitem{cardoso1}V. Cardoso, R Konoplya and P. S. Lemos  {\it Phys. Rev. D} {\bf68},
044024(2003).
\bibitem{wang3}B. Wang, C. Molina, and E. Abdalla, {\it Phys. Rev. D} {\bf63}, 084001(2000).
\bibitem{cardoso2}V. Cardoso and P. S. Lemos    {\it Phys. Rev. D}
{\bf64}, 084017(2001).
\bibitem{berti2}E. Berti, V. Cardoso and A. O. Starinets, {\it Class. Quantum Grav.} 26
163001(2009).
\bibitem{price1}R. H. Price, {\it Phys. Rev. D} {\bf5}, 2419(1972).


\bibitem{ding}C. Ding. S. Chen and J. Jing,{\it Phys. Rev. D} \textbf{81}
024028(2009).
\bibitem{majhi}B. R. Majhi, {\it Phys. Lett. B} \textbf{686} 49(2010).
\bibitem{setare}M. R. Setare and D. Momeni, arXiv:1002.0185v1 [hep-th].
\bibitem{songbai}S. Chen and J. Jing, {\it Phys. Lett. B} \textbf{687}
124(2010).
\bibitem{konoplyaqnm}Konoplya R. {\it A. Phys. Lett. B} \textbf{679}
499(2009).
\bibitem{wang}C. Wang and Y. Gui,  {\it Astrophys and Space Sci} \textbf{325},
85(2010).
\bibitem{lin}K. Lin, N. Yang and J. Li, {\it Int. J. Theor. Phys} {\bf 50} 48(2010).
\bibitem{nv}N. Varghese and V C Kuriakose, arXiv:1011.6608[gr-qc].
\bibitem{gundlach}C. Gundlach, R. H. Price, and J. Pullin, {\it Phys. Rev. D} {\bf49}, 883
(1994).
\bibitem{zhidenko}R. A. Konoplya and A. Zhidenko, {\it Phys. Rev. D}
\textbf{77} 104004(2008).
\bibitem{price2}R. H. Price, {\it Phys. Rev. D} {\bf5}, 2439(1972).

\bibitem{ruffini}R. Ruffini, in Black Holes: les Astres Occlus (Gordon and Breach, New York, 1973).
\bibitem{birrell} N. D. Birrell and P. C. Davies, {\it Quantum Fields In Curved Space}, 2nd edn. (Cambridge, Uk:
Univ. Pr. 1982).

\end{thebibliography}
\end{document}